# A new thin film photochromic material: Oxygen-containing yttrium hydride

Trygve Mongstad[1,*], Charlotte Platzer-Björkman[2], Jan Petter Maehlen[1], Lennard P. A. Mooij[3], Yevheniy Pivak[3], Bernard Dam[3], Erik S. Marstein[1], Bjørn C. Hauback[1] and Smagul Karazhanov[1]

[1] Institute for Energy Technology, Box 40, NO-2027 Kjeller, Norway
[2] Uppsala University, Solid State Electronics, Box 534, SE-751 21 Uppsala
[3] Delft University of Technology, Materials for Energy Conversion and Storage, Julianalaan 136, 2628BL Delft, The Netherlands

Abstract:

In this work we report on photochromism in transparent thin film samples of oxygen-containing yttrium hydride. Exposure to visible and ultraviolet (UV) light at moderate intensity triggers a decrease in the optical transmission of visible and infrared (IR) light. The photo-darkening is colour-neutral. We show that the optical transmission of samples of 500 nm thickness can be reduced by up to 50% after one hour of illumination with light of moderate intensity. The reaction is reversible and samples that are left in the dark return to the initial transparent state. The relaxation time in the dark depends on the temperature of the sample and the duration of the light exposure. The photochromic reaction takes place under ambient conditions in the as-deposited state of the thin-film samples.

Graphical abstract:

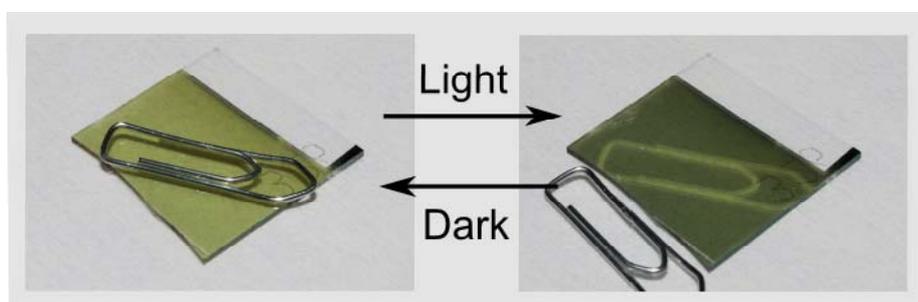

Research highlights:
- Photochromic response in the as-deposited state at ambient conditions
- Sensitive to visible and ultraviolet light
- Color-neutral photo-darkening
- Persistent photoconductivity accompanies the photochromic darkening
- Reversible reaction, material relaxes back to transparent state when left in dark

---

[*] Corresponding author. E-mail adress: trygve.mongstad@ife.no

# 1. Introduction

Photochromic materials constitute an intriguing subset of solar energy materials which respond to electromagnetic radiation by reversibly changing their colour [1]. This unique property is relevant for a broad range of applications, e. g. in smart windows [2], displays [3] and high-density optical memories [4]. To date, photochromism has been most thoroughly investigated in organic materials [5]. The inorganic photochromic compounds have received less attention and the number of inorganic materials known to exhibit photochromism is limited. The most investigated thin-film inorganic photochromic materials are the transition metal oxides $WO_3$ [6] and $MoO_3$ [7], which have also been explored for application in electrochromic devices [8] that change optical properties in response to an applied external voltage.

The inorganic yttrium-hydrogen system has over the last 15 years been thoroughly investigated because of the fascinating reversible optical transition that accompanies hydrogen uptake in yttrium films [9]. Yttrium metal films can absorb hydrogen and become yellowish transparent when the wide band gap semiconductor yttrium trihydride ($YH_3$, $E_g$ = 2.6 eV [10]) is formed. Yttrium hydride and the similar rare earth hydrides have been investigated for applications in smart windows, and an all-solid-state electrochromic device based on the phenomenon has been demonstrated [11]. Photochromism and persistent photoconductivity in transparent yttrium hydride have been reported earlier at extremely low temperatures [12] and at extreme $H_2$ pressure (5.8 GPa) and high light intensity [13], but for yttrium hydride the current report is the first report of these phenomena at room temperature, ambient pressure and moderate light intensity. The films prepared in this work were found to hold a substantial amount of oxygen, which may play an important role for the photochromism of yttrium hydride.

# 2. Experimental

The photochromic thin films of transparent oxygen-containing yttrium hydride were prepared by reactive magnetron sputtering from a metallic yttrium target in an atmosphere of argon and hydrogen in a Leybold Optics A550V7 in-line sputtering system. A metallic yttrium target (99.99%) was used with argon (purity 5N) and hydrogen (6N) gas in the chamber during deposition. The base pressure of the chamber was $10^{-4}$ Pa. The depositions were performed at a pressure of 0.4 Pa, with a 4:1 gas flow ratio of Ar to $H_2$. Films were deposited on glass, carbon and silicon substrates. The thickness of the samples was measured by a stylus profilometer.

The optical transmission and reflection of the films were measured by Ocean Optics QE65000 and NIRQUEST optical spectrometers. The photochromic darkening of the samples was monitored by measuring the transmission with the optical spectrometers while the film was exposed to a stronger light source incident at a different angle. The exposure was done by illumination from a solar simulator (AM 1.5) with an intensity of 0.1 W/cm$^2$, while the measuring probe light from the spectrometer was several orders of magnitude weaker, enabling us to also monitor the transmission under close to dark conditions. Simultaneous resistivity and transmission measurements were done under exposure, where the resistivity was monitored by measuring the current through a thin slab of the sample contacted by two thin aluminium pads deposited on the glass prior to the sample. The resistivity response was confirmed by ex-situ measurements of the resistivity by four-point-probe measurements.

Structural characterization of the samples by X-ray diffraction and a detailed comparison of the optical and electrical properties with literature for yttrium hydride films is available in earlier published work [14]. Neutron reflectometry (NR) and Rutherford backscattering spectrometry (RBS) was carried out to investigate the chemical composition and formation of surface oxide layers. NR measurements were carried out at the EROS time-of-flight reflectometer at Laboratoire Leon-Brillouin, Saclay, France. RBS measurements were performed at the Tandem Accelerator Laboratory, Uppsala University, Sweden.

## 3. Results and discussion

3.1 Optical and electrical behaviour

Fig. 1 displays photographs of a thin film sample before and after one hour of exposure to AM1.5 sunlight from a solar simulator. The initial state is yellowish transparent because of absorption of blue and violet light due to band-to-band excitations, while the photochromic optical density change is colour-neutral. The photo-darkening in Fig. 1 occurred at 0.1 W/cm$^2$ illumination at ambient conditions in air and in the as-deposited state of the films without any chemical pre-treatment. The change in optical density is accompanied by a substantial reduction in the resistivity of the material. When left in dark, the films relax (bleach) back to the initial transparent state. Images imprinted such as the shadow of the paper-clip in Fig. 1 remain well-defined until the sample bleach back to its initial optical state.

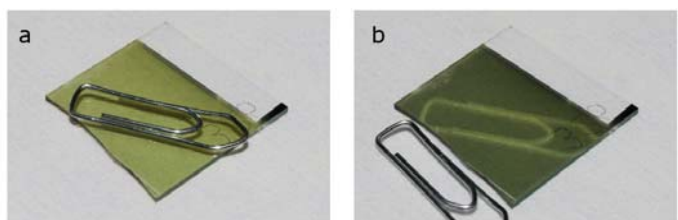

*Fig. 1. Visual appearance of a 550 nm film of oxygen-containing yttrium hydride deposited by reactive magnetron sputtering on a glass substrate. (a) Before light exposure and (b) after one hour of exposure to solar radiation. A paperclip is used to emphasize the contrast between the yellowish-transparent unexposed state and the photo-darkened state.*

The optical transmission and reflection spectra measured of the initial and the photo-darkened state of a thin film sample of oxygen-containing yttrium hydride are displayed in Fig. 2. The transmission decreased over the whole wavelength range from the band gap and into the near IR region. For the data displayed in Fig. 2(a), the average transmission in the wavelength range 500 – 900 nm was reduced by 49 % after 1 hour of illumination of intensity 0.1 W/cm$^2$ in a solar simulator (AM 1.5). Fig. 2(b) displays the reflection, which was also reduced as a result of the photochromic reaction. The optical density displayed in Fig. 2(c) is calculated as $D(\lambda) = \log(T(\lambda)^{-1})$, were $T(\lambda)$ is the optical transmission. The optical density shows a significant change from the band gap and over the whole visible and near IR range, but the effect gradually weakens for wavelengths above 1200 nm. The band gap of the sample was not changed, and the small displacements of the interference fringes demonstrate a very small change in the real part of the refractive index. The absorption coefficient in the inset of Fig. 2(c), calculated as $\alpha = \ln((1-R(\lambda))/T(\lambda))$, shows the same trend as the optical density, but the omits the interference effects.

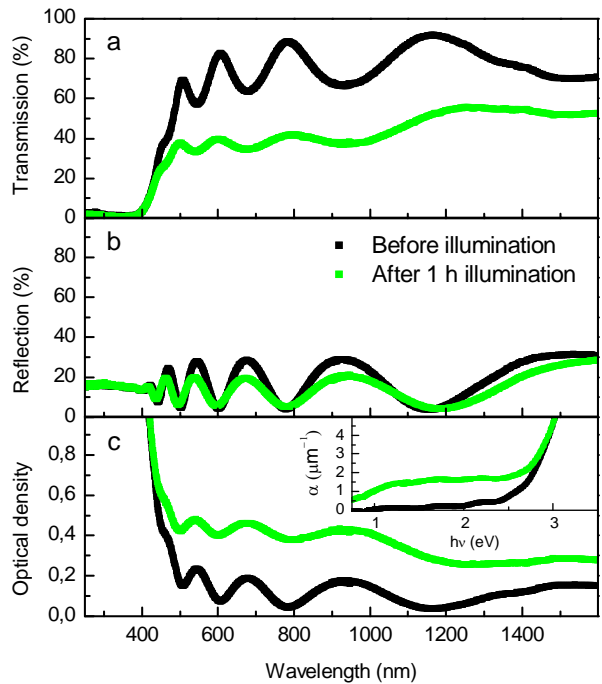

*Fig. 2. Optical spectra in the UV to near IR range for a 520 nm thick oxygen-containing yttrium hydride film on glass substrate before (black) and after (green) one hour of exposure to solar radiation. (a) Transmission spectra, (b) reflection spectra and (c) calculated optical density, with inset that shows the calculated absorption coefficient before and after illumination for the presented data. The oscillations in the data in (a-c) are due to thin-film optical interference.*

Fig. 3 shows the response in optical transmission and electrical resistivity of a sample of oxygen-containing yttrium hydride that was subject to several light exposure cycles. In Fig. 3(a), the initial transmission of the current sample before excitation was 76% in the wavelength area 500-900 nm, and after several excitations of up to 30 minutes of sunlight the transmission was reduced to 54%. The reaction was reversible, and the bleaching relaxation back to the initial transparent state takes place when the sample is left in dark, although at a slower rate than the activating reaction. The sample did not relax completely back to the initial state between each excitation, which explains the general decreasing trend in the transmission and the resistivity. As seen in Fig. 3(b), the sample exhibited persistent photoconductivity with a photo-response similar to the optical photo-response. The resistance measured at the initial state corresponds to resistivity in the order of $10^5$ Ωcm [14], and was reduced by a factor of 20 during the demonstrated illumination sequence. Several hours of light exposure can reduce the resistivity by a factor of up to 100.

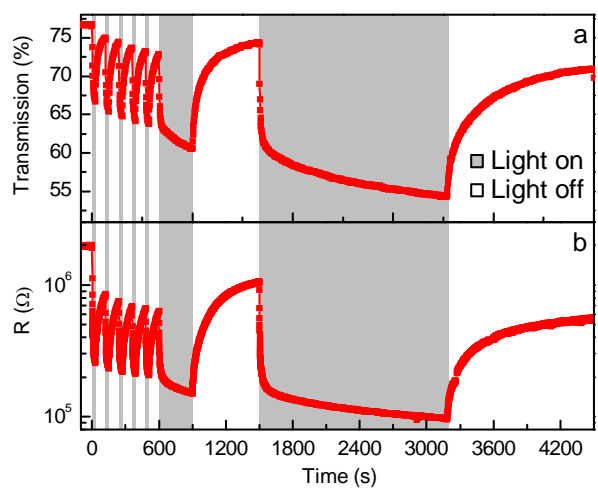

*Fig. 3. (a) Optical transmission and (b) electrical resistance as a response to AM1.5 solar illumination for a 500 nm thick sample. The figure shows five short excitations of 30 seconds, followed by one excitation of 5 minutes and one of 30 minutes. The transmission is an average value for light with wavelength in the range 500 - 900 nm.*

The wavelength of the excitation light had an effect on the strength of the photochromic reaction. Blue and UV light with photon energies above the band gap of the material (2.6 eV) gave the strongest photo-response. Experiments with blue (2.70 eV), green (2.33 eV) and red (1.99 eV) diode illumination of similar intensity showed that also red and green light gave a change in the resistivity, however the magnitude of the change was strongest for the blue, lower for the green and substantially lower for the red. The fact that illumination with photon energies above the band gap energy gave a stronger reaction suggests that excitation of charge carriers from the valence band is essential to obtain a photochromic response.

A memory effect is present in the photochromic response, which results in faster photo-darkening of areas that have been exposed to light before. This is clearly demonstrated by observing subsequent illumination cycles of imprinted images like that of the paperclip in Fig. 1. The image can completely disappear under bleaching in dark, but will become visible again upon a new exposure to light. This memory effect can persist for weeks. A similar memory effect has been observed in photochromic transition metal oxides [7,15].

3.2 Synthesis and composition

Reactive sputtering is an unconventional way of preparing thin-film metal hydrides. Yttrium hydride films are most commonly prepared by hydrogenation of a palladium-capped thin film of pure yttrium metal, where the palladium both serves as a protection against oxidation and as a catalysing layer for the hydrogen uptake [16]. In this work, even though we did not incorporate oxygen intentionally, the reactively sputtered films were found to hold a substantial amount of oxygen. RBS and NR measurements estimated the oxygen content to be between 8 and 32 at% in samples exhibiting photochromism. Samples with higher oxygen content were also photochromic, but the response of these samples was lower. Lower oxygen content was not achieved by reactive sputter deposition. A residual gas analyser showed that water vapour was present in the deposition chamber during the reactive deposition process. Yttrium is highly reactive to oxygen, and oxygen from water vapour can easily be incorporated in the samples during deposition. An isotropic distribution of oxygen through the thickness of the sample found by RBS and NR is another indication that oxygen is incorporated into the samples during the synthesis. Some post-deposition oxidation is unavoidable, since the samples were not protected against oxidation by any capping layer. A surface layer of the sample is oxidized, and this layer protects the rest of the film from oxidation and hydrogen desorption. The samples were stable in air; they did not completely oxidize and retained the photochromic sensitivity to light after storage in air for at least one year.

3.3 Photochromic mechanism in oxygen-containing yttrium hydride

At present the mechanism of the observed photochromism remains unresolved. It is well known that yttrium hydride and other rare earth hydrides can change optical properties dramatically upon hydrogen absorption, even without going through a structural phase transition [10]. We have previously shown that reactive sputtering of yttrium can create two structurally similar oxygen-containing hydrides with very different optical properties [14]. While transparent $YH_3$ is known to exhibit an hcp crystal lattice, the reactively sputtered samples had an expanded $YH_2$-like fcc lattice, probably stabilized by oxygen impurities. Both transparent and black yttrium hydrides were formed in the reactive sputtering process, and the only structural difference found was a change in the lattice parameter of the fcc lattice. The lattice parameters for transparent and black hydrides were 5.35 and 5.26 Å, respectively. Thus, knowing that a small structural change can result in a large difference in the optical properties, a light-induced localized structural transition seems a reasonable explanation for the photochromic optical changes observed. However, an attempt to model the optical properties using a Bruggeman effective medium approximation with a transparent yttrium hydride matrix and metallic yttrium hydride particles could not account for the observed changes in the optical spectra. Further investigations of the optical absorption mechanism should involve a study of the valence state of the metal atoms, detailed structural analysis and comparisons with the different optical states of yttrium hydride and other rare earth hydrides.

The importance of oxygen for the photochromic effect seems evident. Photochromism in yttrium hydride at room temperature has so far only been observed in oxygen-containing samples. However, intriguingly, the samples with the lowest oxygen concentration appeared to have a quicker and stronger photochromic response under our test conditions with solar light. We also observed an increase in the resistivity and in the band gap with increasing oxygen content, similarly to what was found for $GdO_yH_x$ films in the work of Miniotas et al. [17]. The increase in band gap may be partly or entirely responsible for the reduction in the photochromic response with increasing oxygen content. We have insufficient information to evaluate the actual role oxygen plays in the photochromism and which level of oxygen concentration gives the strongest photochromic effect.

In transition metal oxides photochromism is generally explained by a model of ion insertion. Hydrogen ions from water at the surface or interior is dissociated by photo-exited positive charge carriers (holes) [7]. The hydrogen ions can diffuse in the oxide and change the valence state of the metal ions by formation of transition metal bronzes. The optical absorption mechanism in transition metal oxides is subject to dispute. Intervalence charge transfer and small polaron absorption [18] are the mechanisms with most support at present [19]. The optical density change and the reaction rate are similar in oxygen-containing yttrium hydride and in transition metal oxides, but the materials differ significantly in several other aspects. Firstly, the reaction is reversible by self-relaxation in darkness for the oxygen-containing yttrium hydride, while transition metal oxides must be bleached electrochemically. Another way to bleach transition metal oxides is by heat treatment in an oxygen-containing atmosphere. We did indeed observe that the bleaching was accelerated by moderate heating to 50 ºC, but heating was not necessary to bleach the photo-darkened material. Secondly, the oxygen-containing yttrium hydride is sensitive to visible light, whereas transition metal oxides in their pure form only react to UV light [3,20]. This may be explained by the smaller band gap of transparent yttrium hydride that allows electrons to be excited by blue and violet light. Thirdly, the darkening of the yttrium hydride is colour-neutral, whereas the transition metal oxides obtain a blue colour under illumination, which may imply that the light-absorbing mechanism for oxygen-containing yttrium hydride differs from the explanation given for the transition metal oxides.

Yttrium hydride is considered a member of the rare earth hydrides [21], which all exhibit very similar physical properties. The finding of the photochromic effect in oxygen-containing yttrium hydride could provide an opening to a new class of inorganic photochromic compounds. Erbium hydride has also been reported to form an fcc as opposed to hcp lattice when prepared by reactive sputtering [22], and it is probable that this kind of material also exhibits photochromism. Further investigation should be carried out to explore photochromism in other oxygen-containing rare earth hydrides and to enhance the reaction rate and strength of the photochromism in oxygen-containing yttrium hydride.

## 4. Conclusion

We report a photochromic effect observed in oxygen-containing yttrium hydride thin films. The effect is reversible and the photo-darkened films bleach back to the initial transparent state when left in dark. The photochromic activity at ambient conditions and the sensitivity to visible light makes this material interesting for further investigations for employment in technological applications as smart windows and optical memory. The persistent photoconductivity that lowers the resistivity by up to a factor 100 by illumination may also prove interesting for applications in memory devices and optical sensors.

## Acknowledgements

The NANOMAT program of the Research Council of Norway is acknowledged for financial support. We would like to thank A. Holt, H. Schreuders, F. Cousin and G. Possnert for interesting discussions and help with the experimental work.